\def\year{2020}\relax
\documentclass[letterpaper]{article} 
\usepackage{aaai20}  
\usepackage{times}  
\usepackage{helvet} 
\usepackage{courier}  
\usepackage[hyphens]{url}  
\usepackage{graphicx} 
\usepackage{graphicx}  
\usepackage{amsopn}
\usepackage{amsfonts}
\usepackage{textcomp}
\usepackage{pgfplots}
\usepackage{boldline}

\usepackage{tikz}
\usetikzlibrary{matrix, positioning, arrows, shapes.geometric}
\pgfdeclarelayer{background}
\pgfsetlayers{background,main}
\tikzstyle{box}=[draw, fill=white!20, text width=5em, 
    text centered, minimum height=2.5em]
\tikzstyle{blank}=[text width=5em, 
    text centered, minimum height=2.5em]
\tikzstyle{ar} = [->, shorten <=1pt, shorten >=1pt, >=stealth]

\title{Privacy-Preserving Gaussian Process Regression -- A Modular Approach to the Application of Homomorphic Encryption}
\author{Peter Fenner and Edward O. Pyzer-Knapp\\ 
IBM Research UK\\
  Sci-Tech Daresbury\\
  Warrington, UK.\\
peter.fenner1@ibm.com, epyzerk3@uk.ibm.com 
}
\begin{document}

\maketitle

\begin{abstract}
Much of machine learning relies on the use of large amounts of data to train models to make predictions. When this data comes from multiple sources, for example when evaluation of data against a machine learning model is offered as a service, there can be privacy issues and legal concerns over the sharing of data. Fully homomorphic encryption (FHE) allows data to be computed on whilst encrypted, which can provide a solution to the problem of data privacy. However, FHE is both slow and restrictive, so existing algorithms must be manipulated to make them work efficiently under the FHE paradigm. Some commonly used machine learning algorithms, such as Gaussian process regression, are poorly suited to FHE and cannot be manipulated to work both efficiently and accurately. In this paper, we show that a modular approach, which applies FHE to only the sensitive steps of a workflow that need protection, allows one party to make predictions on their data using a Gaussian process regression model built from another party's data, without either party gaining access to the other's data, in a way which is both accurate and efficient. This construction is, to our knowledge, the first example of an effectively encrypted Gaussian process.
\end{abstract}

\section{Introduction}

As increasingly large amounts of data are becoming available, machine learning techniques are increasingly essential for gaining insights into trends and patterns which can be used to make predictions about new data. Indeed, many industries already benefit greatly from the use of machine learning for tasks such as drug discovery and fraud detection, and thus as the application areas continue to broaden, the servitization of machine learning capabilities seems inevitable. However, this capability relies on the sharing of data for training the required machine learning models, or for predicting on new data, and is thus limited by any constraints on the disclosure of this information.

For many reasons, including legal restrictions or competitive advantage, people and organisations may wish to keep their data private. In the case where machine learning is offered as a service, if the client wishes to keep their data private then they may be unwilling to share it with the service provider. The service provider is unlikely to be willing to share their model or data with the client, as this would undermine their ability to continue offering the service. Since neither party is willing to share their data with the other, combining the data to make predictions poses a problem.

One potential solution to this problem is found in fully homomorphic encryption (FHE), which allows for calculations to be made on encrypted data. Privacy is ensured by encrypting the client's data and sending it to the service provider, then allowing the service provider to perform calculations on the encrypted data before returning the results to the client, who can then decrypt them. This problem has been considered for several machine learning algorithms. The paper \emph{Implementing ML Algorithms with HE} \cite{du_implementing_2017} describes implementations of linear regression and K-means clustering under FHE, and gives an overview of other machine learning algorithms which have been implemented by others, including Ridge Regression \cite{nikolaenko_privacy-preserving_2013}, Linear Means Classifiers \cite{graepel_ml_2012}, Naive Bayes, Decision Trees and Support Vector Machines \cite{bost_machine_2014}, K-Nearest Neighbour \cite{samet_privacy_2007,clifton_privacy-preserving_2003}, and Neural Networks \cite{dowlin_cryptonets:_2016}.

The use of FHE for machine learning is not, however, the panacea it may at first appear to be. Performing computation on encrypted data is much slower than performing the same computation on plain-text, and there are limits on the types of operations which can be used. Some machine learning algorithms will therefore lend themselves to FHE better than others. For example, parts of some popular algorithms cannot realistically be accurately computed under FHE and must be approximated, with more accurate approximations requiring greater computational expense, and so a choice must be made between speed and accuracy.

Gaussian process regression is one example of a machine learning algorithm which provides a challenge when applying FHE, and as a result there does not (to our knowledge) currently exist any application of FHE to Gaussian processes. Smith et al. \cite{smith_2018} have considered an alternative approach to the problem of applying privacy-enhancing techniques to Gaussian processes, using differential privacy, but their approach protects only the target values of the data and not the feature vectors.

In this paper we identify why parts of Gaussian process regression are unsuited for computation under FHE and demonstrate that this problem can be avoided by splitting the algorithm into parts in such a way that only some parts need to be computed homomorphically. This enables the user to circumvent some stages of the algorithm which are not amenable to computation under FHE, and to optimize those which are computed under FHE individually, with the end result being both faster and more accurate than performing the whole algorithm in a single block.

In Section \ref{methods} we review the relevant details of Gaussian process regression and FHE. In Section \ref{fhe4gp} we look at the complications and restrictions of adapting Gaussian process regression to FHE and introduce our proposed solution, which splits the homomorphic computation into smaller components. We also analyse the security of our solution and give details of an implementation. In Section \ref{results} we present our results and show that, compared to an implementation which does not split up the homomorphic computation, our approach is both faster and more accurate.

\section{Methods} \label{methods}

\subsection{Gaussian Processes} \label{methods_gp}

A Gaussian process (GP) is a stochastic process such that every finite set of random variables has a multivariate Gaussian distribution. A Gaussian process can be used to define a probability distribution over the set of continuous functions, and by incorporating training data we can construct a model of the relationship represented by the data. In constructing this model we assume there is an underlying function $f$ and our training data consists of a sample of values in the function's domain (feature vectors), and the corresponding outputs of the function (target vectors), possibly with some error in measurement. A Gaussian process model can be used for regression, to make predictions about the behaviour of the function on feature vectors outside of the sample.

We build the GP model by first placing a prior distribution over functions based only on our beliefs about the general form that the function should take, then updating this distribution using known function values. The prior distribution is Gaussian and has mean value 0 at all points, so the distribution is defined entirely by the covariances of the function values. These covariances are given by a kernel function, $k( x, x' )$. A common class of kernel functions are \emph{isotropic} kernels, which measure the distance $d(x, x')$ with respect to some distance metric $d(\cdot, \cdot)$, then calculate the covariance based on just this distance. In this paper we consider only isotropic kernels.

Although the kernel function gives a covariance for all pairs of points in the function's domain, when building a GP model to make predictions on new feature vectors, we only need to calculate the covariances between each pair of feature vectors in the training data and new data. Given training data $\mathbf{x} = [x_1, x_2, \ldots, x_n]$ and $\mathbf{y} = [y_1, y_2, \ldots, y_n] = [f(x_1), f(x_2), \ldots, f(x_n)]$, and new data  $\mathbf{x}_{*} = [x^{*}_1, x^{*}_2, \ldots, x^{*}_m]$ for which we wish to predict $\mathbf{y}_{*} = [f(x^{*}_1), f(x^{*}_2), \ldots, f(x^{*}_m)]$, we calculate:

\begin{itemize}
\item the $n \times n$ covariance matrix $K$ defined by $k_{i,j} = k(x_i, x_j)$,
\item the $m \times n$ matrix $K_*$ defined by $k^*_{i,j} = k(x^{*}_i, x_j)$,
\item the $m \times m$ matrix $K_{**}$ defined by $k^{**}_{i,j} = k(x^{*}_i, x^{*}_j)$.
\end{itemize}

Then our assumed prior distribution gives us
\begin{displaymath}
    \left[
        \begin{tabular}{c}
            $\mathbf{y}$   \\
            $\mathbf{y}_{*}$  
        \end{tabular}
    \right]
    \sim \mathcal{N}
    \left(
        \mathbf{0},
        \left[
            \begin{tabular}{cc}
                $K$ & $K_{*}^T$ \\
                $K_{*}$ & $K_{**}$
            \end{tabular}
        \right]
    \right).
\end{displaymath}
From this we get that the elements of $\mathbf{y}_{*}$ have Gaussian distributions with means given by $\bar{\mathbf{y}}_{*} = K_{*}K^{-1}\mathbf{y}$ and variances given by $\mathrm{var}(\mathbf{y}_{*}) = \mathrm{diag}(K_{**}) - \mathrm{diag}(K_{*}K^{-1}K_{*}^T)$.

One common isotropic kernel function is the squared exponential function, also commonly known as the radial basis function (RBF) kernel or the Gaussian kernel:
\begin{displaymath}
    k(a, b) = \operatorname{exp}\left(- \frac{d(a, b)^2}{2l^2}\right),
\end{displaymath}
for some choice of lengthscale parameter $l$. In theory, any true metric can be used as the distance measure, but here we focus on the Hamming, Jaccard and Euclidean distance metrics, which are commonly used in machine learning tasks, with the Jaccard being particularly prevalent in cheminformatics. Other metrics could certainly be used in the scheme we propose, but care must be taken to implement them efficiently under homomorphic encryption.

\subsection{Fully Homomorphic Encryption} \label{methods_fhe}

The concept of homomorphic encryption was proposed by Rivest, Adleman, and Dertouzos \cite{Rivest1978}, but the ideas were not realized until 2009 when Gentry designed a functional (but prohibitively slow) fully homomorphic encryption scheme \cite{homenc}. Since then, many homomorphic encryption schemes have been developed, such as the Brakerski-Gentry-Vaikuntanathan (BGV) \cite{brakerski_fully_2011} scheme, and many optimizations have been made. These schemes are homomorphic in addition and multiplication, in that $\mathrm{Encrypt}(x) + \mathrm{Encrypt}(y) = \mathrm{Encrypt}(x+y)$ and $\mathrm{Encrypt}(x)\mathrm{Encrypt}(y) = \mathrm{Encrypt}(xy)$.

Gentry's design begins with a ``somewhat homomorphic" encryption scheme, which behaves homomorphically for a limited number of computations, thus limiting the depth of arithmetic circuits which can be calculated. He then extends this to a ``fully homomorphic" encryption (FHE) scheme with the introduction of bootstrapping, a process which ``refreshes" the ciphertext to allow for more computation. Bootstrapping allows the data to be computed on indefinitely, but is a very slow process.

The BGV scheme allows for more efficient FHE by introducing modulus switching either instead of or in conjunction with bootstrapping. The result is known as leveled homomorphic encryption. In leveled homomorphic encryption, arithmetic circuits of any depth can be computed without bootstrapping, but the depth must be known before setting the encryption scheme's parameters. The work described in this paper uses leveled FHE without bootstrapping.

\subsubsection{HElib}
We make use of the homomorphic encryption library HElib (\citeauthor{helib}), which implements the BGV scheme. The BGV scheme works over polynomial rings of the form $\mathbb{A} = \mathbb{Z}[X]/\Phi_m(X)$, where $\mathbb{Z}[X]$ is the ring of integer-coefficient polynomials in $X$ and $\Phi_m(X)$ is the $m$th cyclotomic polynomial. The plaintext space for the scheme is $\mathbb{A}_{p^r}$, for any prime $p$ and positive integer $r$, where we define $\mathbb{A}_q = \mathbb{A}/q\mathbb{A}$ for any integer $q$. In our implementation of GP regression we set $r=1$, so we shall use this value in the following explanation.

Ciphertexts are vectors in $(\mathbb{A}_q)^2$ for some odd modulus $q$. When using modulus switching we have a sequence of decreasing moduli $q_L \gg q_{L-1} \gg \ldots \gg q_0$, so that ciphertexts lie on $L+1$ different ``levels". Given two moduli $q$ and $q'$, elements of $\mathbb{A}_q$ can be mapped to $\mathbb{A}_{q'}$ by first using the map $[\cdot]_{q}$ which maps an element $x \in \mathbb{A}_q$ to the unique element of $\mathbb{A}$ which is equal to $x$ mod $q$ but has each coefficient in the range $[-q/2, q/2)$. This element of $\mathbb{A}$ can then be interpreted as an element of $\mathbb{A}_{q'}$.

A level-$i$ ciphertext $\mathbf{c} = (\mathfrak{c}_0, \mathfrak{c}_1)$ is said to encrypt a plaintext $\mathfrak{m} \in \mathbb{A}_p$ with respect to a secret key $\mathbf{s} = (1, \mathfrak{s}) \in \mathbb{A}^2$ if $[\langle \mathbf{s}, \mathbf{c} \rangle]_{q_i} = [\mathfrak{c}_0 + \mathfrak{s}\mathfrak{c}_1]_{q_i} = \mathfrak{m} + p\mathfrak{e}$ for an error term $\mathfrak{e}$ which satisfies $p \left\Vert \mathfrak{e} \right\Vert \ll q_i$. Decrypting $\mathbf{c}$ is then a matter of calculating $[[\langle \mathbf{s}, \mathbf{c} \rangle]_{q_i}]_p$. The result of $[\langle \mathbf{s}, \mathbf{c} \rangle]_{q_i}$ is referred to as ``noise", and decryption works because the noise is equal to $\mathfrak{m}$ mod p. However, the maximum possible size of the noise grows with each calculation: the sum of ciphertexts $\mathbf{c}_0$ and $\mathbf{c}_1$ has noise equal to the sum of the noises of $\mathbf{c}_0$ and $\mathbf{c}_1$, and the product $\mathbf{c}_0\mathbf{c}_1$ has noise equal to the product of the noises of $\mathbf{c}_0$ and $\mathbf{c}_1$. Noise can therefore grow very quickly when many sequential multiplications are performed. By definition the noise is guaranteed to have each coefficient in the range $[-q_i/2, q_i/2)$. We think of $q_i/2$ as a ``noise ceiling", and if any coefficient grows too large it may exceed the noise ceiling and overflow, causing decryption to fail.

The sequence of decreasing moduli described above (known as the \emph{modulus chain}) provides a way of managing the noise level. Freshly encrypted ciphertexts are encrypted at the highest level in the chain. It is possible to convert a valid level-$i$ ciphertext $\mathbf{c}$ to a valid level-$(i-1)$ ciphertext which encrypts the same plaintext by scaling $\mathbf{c}$ by $(q_{i-1}/q_i)$ and rounding to the closest level-$(i-1)$ ciphertext $\mathbf{c}'$ such that $\mathbf{c}' = \mathbf{c}$ mod p. This lowers the level of noise by a factor of $(q_{i-1}/q_i)$ but also lowers the noise ceiling by the same factor. This can help to manage the increase in noise resulting from multiplication. When evaluating an arithmetic network homomorphically, HElib will automatically decrease the modulus of a ciphertext when necessary. It will often decrease the modulus level of two ciphertexts before multiplying them, so the required length of modulus chain is correlated with the number of sequential multiplications in the network.

We can make an arbitrarily large number of computations on a piece of data, so long as we know the arithmetic network in advance and create a long enough modulus chain. However, the length of the modulus chain has a significant effect on the speed of computation under BGV. The per-gate computation in BGV without bootstrapping is $\tilde{O}(\lambda \cdot L^3)$, where $\lambda$ is the security parameter and $L$ is the length of the modulus chain \cite{brakerski_fully_2011}. Any change in the depth of an arithmetic network can therefore make a large change in the time needed to evaluate it, so careful planning of algorithms is incredibly important.

\section{FHE for Gaussian Processes}\label{fhe4gp}

\subsection{Technical Considerations}

When working with FHE we are limited in the types of operation we can use. In BGV the only arithmetic operations we can use are addition and multiplication, which means we can calculate only polynomials. The length of the modulus chain, $L$, required for decryption to succeed roughly correlates with the number of sequential multiplications in the algorithm. For a degree $d$ polynomial this is, at best, roughly equal to $\log_2(d)$. Since the per-gate computation time is cubic in $L$ we are in practice limited to performing calculations using just low-degree polynomials, restricting our ability to approximate non-polynomial functions. Many algorithms involve the evaluation of functions which cannot be approximated well by a low degree polynomial. This leaves the user with a choice between using a high degree polynomial, in which case the algorithm will run slowly under FHE, or a lower degree polynomial, resulting in poor accuracy.

Gaussian process regression is an example of an algorithm for which these problems are significant. It is impossible to accurately approximate GP regression using low-degree polynomials, mainly due to the use of a kernel function. Most isotropic kernel functions are largest when the distance is close to zero, and tend to zero as the distance gets large. However, a polynomial will (unless it is constant) be much larger in magnitude at large inputs than at 0. The best polynomial approximation to the squared exponential kernel, which our implementation uses, is a Taylor expansion around 0, but even a large degree Taylor approximation diverges very quickly outside of a relatively small region. The point at which the approximation breaks down is proportional to the lengthscale, so problems which require a small lengthscale relative to the size of the domain will be approximated very badly.

\begin{figure*}
    \centering
	\tikzstyle{stepbox1}=[draw, fill=white!20, text width=4.3em, text centered, minimum height=2.5em, inner sep=0pt]
	\tikzstyle{stepbox2}=[stepbox1, fill=white!40!black, text=white]
	\tikzstyle{databox1}=[stepbox1, rounded corners]
	\tikzstyle{databox2}=[stepbox2, rounded corners]
    \begin{tikzpicture}[every node/.style={font=\small}]
    \node (mean) [databox2] {Mean};
    \path (mean)+(0,-1) node (var) [databox2] {Variance};
    	\path (var)+(-1.9,0) node (ms) [stepbox2] {Matrix\\sum};
    	\path (ms)+(-1.3,1) node (mm) [stepbox2] {Matrix\\products};
    	\path (mm)+(-1.3,1) node (in) [stepbox1] {Invert\\matrix};
    	\path (in)+(-1.9,0) node (co1) [stepbox1] {Apply\\kernel};
    	\path (co1)+(-1.9,0) node (di1) [stepbox1] {Calculate\\distances};
    	\path (mm)+(-3.2,0) node (co2) [stepbox2] {Apply\\kernel};
    	\path (co2)+(-1.9,0) node (di2) [stepbox2] {Calculate\\distances};
    	\path (ms)+(-4.5,0) node (co3) [stepbox2] {Apply\\kernel};
    	\path (co3)+(-1.9,0) node (di3) [stepbox2] {Calculate\\distances};
    	\path (mm)+(-7.1,1.5) node (ta) [databox1] {Target\\vectors};
    	\path (di1)+(-2,-0.5) node (fe1) [databox1] {Feature\\vectors};
    	\path (di3)+(-2,0) node (fe2) [databox2] {Feature\\vectors};
        
        \path [ar, draw, rounded corners] (ta.15) -| (mm.north) ;
        \path [ar] (fe1.15) edge (di1.west);
        \path [ar] (fe1.east) edge (di2.165);
        \path [ar] (fe2.15) edge (di2.195);
        \path [ar] (fe2.east) edge (di3.west);
        \path [ar] (di1.east) edge (co1.west);
        \path [ar] (di2.east) edge (co2.west);
        \path [ar] (di3.east) edge (co3.west);
        \path [ar] (co1.east) edge (in.west);
        \path [ar] (co2.350) edge (mm.190);
        \path [ar] (co3.350) edge (ms.190);
        \path [ar, draw, rounded corners] (in.south) |- (mm.170);
        \path [ar, draw, rounded corners] (mm.south) |- (ms.170);
        \path [ar] (mm.east) edge (mean.west);
        \path [ar] (ms.east) edge (var.west);
                   
        \path (ta) +(-1.6,-0.5) node (tr) [blank] {\normalsize Training\\data};
        \path (fe2) +(-1.6,0) node (ne) [blank] {\normalsize Test\\data};
        \path (mean) +(0,0.9) node (pr) [blank] {\normalsize Prediction};
      
        \begin{pgfonlayer}{background}
            \path (tr.west |- ta.north)+(0.2,0.15) node (a) {};
            \path (fe1.east |- fe1.south)+(0.15,-0.15) node (b) {};
              
            \path[fill=gray!20,rounded corners, draw=black!50, dashed]
                (a) rectangle (b);           
    
            \path (ne.west |- fe2.north)+(0.2,0.15) node (c) {};
            \path (fe2.east |- fe2.south)+(0.15,-0.15) node (d) {};
              
            \path[fill=gray!20,rounded corners, draw=black!50, dashed]
                (c) rectangle (d);

            \path (mean.west |- pr.north)+(-0.15,-0.05) node (e) {};
            \path (mean.east |- var.south)+(0.15,-0.15) node (f) {};
              
            \path[fill=gray!20,rounded corners, draw=black!50, dashed]
                (e) rectangle (f);
        \end{pgfonlayer}
        
    \end{tikzpicture}

    \caption{The flow of Gaussian process regression broken into components. Boxes with rounded corners are input/output data. Boxes with square corners are modules in the calculation. If no intermediate results are returned to the client then all the shaded modules must be performed homomorphically.}
    \label{gp_fig2}
\end{figure*}
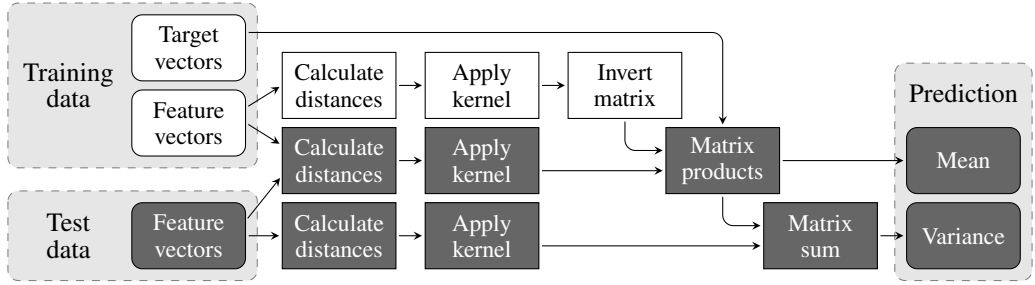

Figure \ref{gp_fig2} shows the Gaussian process algorithm broken up into small modules. As can be seen in the diagram, there are three sequences of modules which can be performed in any order:
\begin{itemize}
\item[a)] calculate the distances between training feature vectors, apply the kernel function to these distances to form a covariance matrix, then invert the covariance matrix,
\item[b)] calculate the distances between training feature vectors and the test feature vectors, then apply the kernel function to these distances,
\item[c)] calculate the distances between test feature vectors, then apply the kernel function to these distances.
\end{itemize}
The results of (a) and (b) are multiplied together with the training target vectors. This gives the mean of the prediction, and another result which is summed together with the result of (c) to get the variance of the prediction.

A naive application of FHE might require the service provider to receive the client's encrypted data, perform the whole algorithm homomorphically, and return an encrypted prediction. With this approach, any module in Figure \ref{gp_fig2} which relies on the client's data, either directly or indirectly, will need to be performed homomorphically. The remaining steps, which rely only on the training data, can be performed in plaintext by the service provider. This approach would need to approximate the entire algorithm as a polynomial, including the kernel function, which cannot be well approximated with a low order polynomial. Achieving an adequate level of accuracy with such an approach would therefore require a very high degree polynomial and as such would be very slow.

However, if we apply FHE with more consideration, for each homomorphic step we can consider whether or not the output from that step needs to be kept secret from the client. If not, then the service provider can return this encrypted output to the client, who can decrypt the output and continue the algorithm in plaintext until the next time the algorithm requires input from the service provider. Analysis of Figure \ref{gp_fig2} reveals that the client only requires the service provider's input in two steps: calculating distances between training feature vectors and test feature vectors, and calculation of the matrix products. Assuming that there is no loss of privacy for the service provider if the client views the output of these steps (we shall revisit this assumption in the next subsection) then these two steps are the only ones which must be performed homomorphically. The resulting workflow (shown in Figure \ref{gp_fig3}) splits up the homomorphic computation and allows us to altogether avoid applying the kernel function homomorphically.

\begin{figure*}
    \centering
	\tikzstyle{stepbox1}=[draw, fill=white!20, text width=4.3em, text centered, minimum height=2.5em, inner sep=0pt]
	\tikzstyle{stepbox2}=[stepbox1, fill=white!40!black, text=white]
	\tikzstyle{databox1}=[stepbox1, rounded corners]
	\tikzstyle{databox2}=[stepbox2, rounded corners]
    \begin{tikzpicture}[every node/.style={font=\small}]
    \node (mean) [databox1] {Mean};
    \path (mean)+(0,-1) node (var) [databox1] {Variance};
    	\path (var)+(-1.9,0) node (ms) [stepbox1] {Matrix\\sum};
    	\path (ms)+(-0.7,1) node (de2) [stepbox2] {Decrypt};
    	\path (de2)+(-0.9,1.3) node (mm) [stepbox2] {Matrix\\products};
    	\path (mm)+(-0.8,1) node (in) [stepbox1] {Invert\\matrix};
    	\path (in)+(-1.9,0) node (co1) [stepbox1] {Apply\\kernel};
    	\path (co1)+(-2.8,0) node (di1) [stepbox1] {Calculate\\distances};
    	\path (mm)+(-0.8,-1.3) node (en2) [stepbox2] {Encrypt};
    	\path (en2)+(-1.9,0) node (co2) [stepbox1] {Apply\\kernel};
    	\path (co2)+(-1.9,0) node (de1) [stepbox2] {Decrypt};
    	\path (de1)+(-0.9,1.3) node (di2) [stepbox2] {Calculate\\distances};
    	\path (di2)+(-0.8,-1.3) node (en1) [stepbox2] {Encrypt};
    	\path (ms)+(-4.3,0) node (co3) [stepbox1] {Apply\\kernel};
    	\path (co3)+(-2.8,0) node (di3) [stepbox1] {Calculate\\distances};
    	\path (mm)+(-8.3,1.5) node (ta) [databox1] {Target\\vectors};
    	\path (di1)+(-2.8,-0.5) node (fe1) [databox1] {Feature\\vectors};
    	\path (di3)+(-2.8,0) node (fe2) [databox1] {Feature\\vectors};
        
        \path [ar, draw, rounded corners] (ta.15) -| (mm.50) ;
        \path [ar] (fe1.18) edge (di1.198);
        \path [ar] (fe1.342) edge (di2.162);
        \path [ar, draw, rounded corners] (fe2.60) |- (en1.west);
        \path [ar] (fe2.east) edge (di3.west);
        \path [ar] (di1.east) edge (co1.west);
        \path [ar, draw, rounded corners] (en1.130) |- (di2.190);
        \path [ar, draw, rounded corners] (di2.east) -| (de1.50) ;
        \path [ar] (de1.east) edge (co2.west);
        \path [ar] (di3.east) edge (co3.west);
        \path [ar] (co1.east) edge (in.west);
        \path [ar] (co2.east) edge (en2.west);
        \path [ar, draw, rounded corners] (en2.130) |- (mm.190);
        \path [ar] (co3.350) edge (ms.190);
        \path [ar, draw, rounded corners] (in.230) |- (mm.170);
        \path [ar, draw, rounded corners] (mm.east) -| (de2.50);
        \path [ar, draw, rounded corners] (de2.230) |- (ms.170);
        \path [ar] (de2.east) edge (mean.west);
        \path [ar] (ms.east) edge (var.west);
                   
        \path (ta) +(-0.4,-1.75) node (tr) [blank] {\normalsize Service~provider};
        \path (fe2) +(-0.7,1.25) node (ne) [blank] {\normalsize Client};

        \begin{pgfonlayer}{background}
            \path (tr.west |- ta.north)+(-0.1,0.12) node (a) {};
            \path (mean.east |- mm.south)+(0.15,-0.12) node (b) {};
              
            \path[fill=gray!20,rounded corners, draw=black!50, dashed]
                (a) rectangle (b);           
    
            \path (ne.west |- mean.north)+(0.2,0.12) node (c) {};
            \path (mean.east |- fe2.south)+(0.15,-0.12) node (d) {};
              
            \path[fill=gray!20,rounded corners, draw=black!50, dashed]
                (c) rectangle (d);

        \end{pgfonlayer}
        
    \end{tikzpicture}

    \caption{Our workflow for encrypted Gaussian process regression. Shaded boxes are steps which involve encrypted data. All other steps are performed on plaintext.}
    \label{gp_fig3}
\end{figure*}

Splitting the algorithm in this way offers significant improvements for speed and accuracy. Each of the two homomorphic steps can now be approximated with its own polynomial, which must necessarily have a smaller degree than that of a polynomial which approximates the whole algorithm to the same level of accuracy. The individual steps can also be optimised separately for FHE, in terms of encoding and parameters, rather than finding a single encoding scheme and set of parameters for the whole algorithm. Most significantly, the kernel function can be calculated in plaintext, which allows us to avoid a slow and inexact encrypted polynomial approximation. We will demonstrate how significant these improvements are in Section \ref{results}.

\subsection{Security analysis}
One downside to this approach is that it requires more interaction between parties, which requires the parties to be in more regular contact with each other. This could be a problem when, for example, the client is in a remote location with poor quality, or noisy, communication channels, but more importantly it also has implications for the level of security.

Although the approach does not reveal any information which directly identifies the protected data, it does give the client access to the transformed data at a couple of intermediate steps of the algorithm, which could be analysed to gain information about the data. It is worthwhile to note that this security problem applies to any implementation of encrypted Gaussian processes, since the predictions made by the algorithm can be used for the same analysis. Returning intermediate results to the client does not introduce the problem, but does potentially worsen it by making more data available for analysis. However there are measures that can be taken to significantly mitigate this.

To provide an example of the difficultly of performing such an attack both with and without the intermediate data, we consider Gaussian process regression over $d$-dimensional binary feature vectors, with $n$ points of training data.

We first consider the case where the attacker has access only to the predictions made by the Gaussian process. In this case an attacker can use the prediction variance to locate training data. Since the predictive model is more confident close to the training data, the training data creates local minima in the prediction variance. By using optimisation methods such as gradient descent, the attacker could find some of these local minima and thus identify some of the training data. In this way, the attacker could find up to $k$ training data points by making $kt$ predictions, where $t$ is the number of iterations that they run their optimisation algorithm for. It is possible that several of these optimisations will find the same local minimum, so $k$ is merely a maximum on the number of data points identified. Even though the attacker can identify some training data, without access to any of the intermediate data they cannot know how many data points are contained in the training data, and thus cannot be sure that they have found all the data.

We now consider the case where the attacker does have access to intermediate data. We start by noting that the second of the two homomorphic steps gives the attacker no extra information to work with. In this step two things are returned to the client: the means of the predicted values and the matrix product $\mathrm{diag}(K_* K^{-1} K_*^T)$. The first of these is part of the prediction that the client wants. The second output differs from the variances of the predicted values only by taking its negative and adding $\mathrm{diag}(K_{**})$, which is an entirely reversible operation. Therefore, giving the output of this homomorphic step to the client gives them no more information than giving them the prediction directly.

We therefore only need to consider the first of the two homomorphic steps. This step gives the client access to the distances between the training feature vectors and their own feature vectors, which gives them some information about the training feature vectors. If no measures are taken to obscure this extra information, an attacker can use a simple attack to find the training data: Predict on each of the unit vectors $\mathbf{e}_i$ and the zero vector $\mathbf{0}$, then each training feature vector has a 1 in the $i$th position if and only if its distance from $\mathbf{e}_i$ is smaller than its distance from $\mathbf{0}$. Once all training feature vectors have been identified in this way, the attacker simply needs to predict on these values to find an approximation of the target values (this may not return the exact target values, depending on the model used, but it should be close). This attack requires only $d + n + 1$ predictions.

There are several measures that can be taken to obscure the distances, thus increasing the difficulty of an attack. For example:
\begin{itemize}
\item Adding a small amount of noise to each distance will prevent the client from knowing the exact distances.  This will affect the prediction, but is a commonly used procedure in Gaussian process regression, and has been shown to improve the numerical performance of the method. 
\item By adding fake data points to the training data before calculating the distances, the service provider can hide the number of training points from the client.
\item The service provider can randomise the order of the distances before sending them to the client, then later put the covariances back in the correct order. This will prevent the client from knowing which distance corresponds to each piece of training data.
\item If the algorithm is modified so the first homomorphic step returns the square of the distance (we recommend this for Euclidean distances, and it can be done with little sacrifice to speed in other cases), the service provider can choose a random integer $c$ for each distance and add this number to the squared distance before sending it to the client. They can later multiply the corresponding covariance by $e^{c/{2l^2}}$ to cancel out this change, again with little sacrifice to speed. In this way the service provider can completely obscure the distances from the client for a small sacrifice in speed.
\end{itemize}

These precautions considerably increase the difficulty in learning the distances and the number of training points, and therefore significantly increase the number of predictions necessary for an attack.

\subsection{Implementation}\label{impl}

The first of the two homomorphic steps is the calculation of the distances between training feature vectors and test feature vectors. The implementation is specific to the choice of distance metric, but we demonstrate that it is possible to implement efficiently for some common distance metrics. We have implemented Hamming distances, Jaccard distances and, with a small modification to the workflow, Euclidean distances.

\paragraph{Hamming distance}
The Hamming distance between two vectors is the number of positions in which they differ from each other. This is commonly used for binary vectors, for example in chemoinformatics where MACCS keys are a common way of describing molecules as 167 dimensional binary vectors. For calculation of Hamming distances we use $\mathbb{Z}_2$ to encode non-negative integers as binary digits. We can then add integers through the appropriate addition and multiplication of the digits. We encode feature vectors in $\mathbb{Z}_2^{n \times 1}$. We regard this as $n$ binary vectors of length 1, each representing an integer stored with a single binary digit. Since $1+1=0$ in $\mathbb{Z}_2$, for feature vectors  $\mathbf{x}, \mathbf{x'} \in \mathbb{Z}_2^{n \times 1}$, the sum $x_{i,1} + x'_{i,1}$ equals 1 if and only if $x_{i,1}$ and $x'_{i,1}$ differ. The sum $\mathbf{x} + \mathbf{x'} \in \mathbb{Z}_2^{n \times 1}$ is therefore equal to 1 in each position where $\mathbf{x}$ and $\mathbf{x'}$ differ, and we get the Hamming distance by adding together these 1s. This is done in steps, where in each step we pair up integers which are stored using $k$ binary digits and add them together, resulting in integers stored using $k+1$ binary digits. After the first step we are left with an element of $\mathbb{Z}_2^{\lceil n/2 \rceil \times 2}$, which is regarded as $\lceil n/2 \rceil$ binary vectors each representing a 2-digit binary number. After the second step we have an element of $\mathbb{Z}_2^{\lceil n/4 \rceil \times 3}$, regarded as $\lceil n/4 \rceil$ binary vectors each representing a 3-digit binary number, and so on until we get an element of $\mathbb{Z}_2^{1 \times \lfloor \log_2(n)+1 \rfloor}$, containing the binary digits of the Hamming distance.

When running this distance algorithm on a single pair of feature vectors, each element of $\mathbb{Z}_2$ is encrypted into a separate ciphertext. By using ciphertext packing \cite{smart_fully_2014} we can store many elements of $\mathbb{Z}_2$ in the same ciphertext and compute on them simultaneously, which can be used to run the distance algorithm on many pairs of feature vectors at no extra computational cost. We use the same packing strategy for Jaccard distances.

\paragraph{Jaccard distance}
The Jaccard distance between two finite sets $A$ and $B$ is $\frac{|A \cup B| - |A \cap B|}{|A \cup B|}$. Binary vectors can be regarded as subsets of some larger finite set ($x_i = 1$ if and only if the $i$th element is present in the subset) so we can take Jaccard distances between binary vectors. This is also commonly used in chemoinformatics. For this calculation we encode feature vectors in $\mathbb{Z}_2^{n \times 1}$ in the same way as for Hamming distance. As the distance is a fraction, and we cannot perform division on encrypted data, we return the fraction as a pair of integers - numerator and denominator - for the client to divide. The denominator is the size of the union, which is the number of indices where one or both vectors are equal to 1. The sum $x_{i,1} + x'_{i,1} - x_{i,1}x'_{i,1}$ is equal to 1 if and only if one or both of $x_{i,1}$ and $x'_{i,1}$ is equal to 1, so we calculate the denominator by calculating $\mathbf{x} +\mathbf{x'} - \mathbf{x} \mathbf{x'}$ and summing the 1s in the same way as for the Hamming distance. The size of the intersection is the number of indices where both vectors are equal to 1. Thus the numerator is the number of indices where exactly one vector is equal to 1, which is precisely the Hamming distance.

\paragraph{Euclidean distance}

The Euclidean distance between vectors $\mathbf{x}$ and $\mathbf{y}$ is $\sqrt{\sum_{i=0}^n (x_i - y_i)^2}$. This involves taking the square root, which we cannot reasonably do on encrypted data. This can be worked around by homomorphically calculating the square of the Euclidean distance and letting the client take the square root. This does not give any extra information to the client since there is a one-to-one mapping between the (positive) distances and their squares. We therefore only need to calculate $\sum_{i=0}^n (x_i - y_i)^2$, which is a very low degree polynomial so is perfectly suited for FHE. In this case we run into another issue. The BGV scheme only works on elements of $\mathbb{Z}_p$ for some prime $p$, whereas the Euclidean distance is defined over all real numbers. We work around this using the same method as \cite{dowlin_cryptonets:_2016}. We scale real numbers by some large scaling factor then round to the nearest integer. Since the integers obtained are potentially very large we use the Chinese Remainder Theorem to split them into vectors of smaller integers, each modulo some prime base. We use primes roughly equal to 10000 and choose a suitably large number of primes to store the scaled numbers. Each number is therefore encoded as a sequence of integers, each an element of $\mathbb{Z}_{p_i}$ for some sequence of plaintext moduli $\mathbf{p}$.

When running this algorithm on a single pair of feature vectors, each element of $\mathbb{Z}_{p_i}$ is encrypted into a separate ciphertext. Again, using ciphertext packing we can perform several instances of the algorithm in parallel. In this case the number of elements we can store in a single ciphertext varies with $p_i$, so, to compute a fixed number of distances, the number of ciphertexts required for each $p_i$ also varies. We use the same packing strategy for multiplying matrices, allowing us to make several predictions at once.

\paragraph{Multiplying matrices}

The second homomorphic step involves multiplying matrices of real numbers. In particular, this step calculates $K_* K^{-1} \mathbf{y}$ and $\mathrm{diag}(K_* K^{-1} K_*^T)$, which requires the evaluation of polynomials of degree 3. This is also a very low degree polynomial with real valued inputs, so is encoded in the same way as for Euclidean distance.

\section{Results}\label{results}

We apply our modular approach to a data set which is a realistic representation of the kind of data for which this technique would provide significant real world value. This data consists of hits from Plasmodium falciparum (P. falciparum) whole cell screening as released through the Medicines for Malaria Venture website (\citeauthor{noauthor_malaria_nodate}) and originates from the GlaxoSmithKline Tres Cantos Antimalarial Set (TCAMS), Novartis-GNF Malaria Box Data  set and St. Jude Children's Research Hospital’s Dataset, and consists of molecules described using MACCS keys \cite{maccs} - a common cheminformatics descriptor, generated using the RDKit software (\citeauthor{landrum_rdkit:_nodate}), consisting of 167-dimensional binary feature vectors. Pharmaceutical companies are unlikely to want to expose any intellectual property during the drug-discovery process, but may want to use machine learning models built on external data to accelerate the discovery process. The owner of this data, if it is not openly available, is unlikely to want to expose their IP either, given the potential value it has for the user.

We compare to a non-modular approach which does not return anything to the client until the (encrypted) prediction has been calculated. All steps up to this point which rely on the client's data must therefore be performed homomorphically. Steps which do not rely on the client's data can still be performed in plaintext by the service provider (see Figure \ref{gp_fig2}). Since the results are real numbers, we use the same technique as for multiplying matrices: multiply by a large scaling factor, round to the nearest integer, and split the large integer into several smaller ones using the CRT.

Since the kernel cannot be calculated exactly we approximate it with a degree 6 Taylor expansion about 0, which provides a good approximation for distances which are small relative to the lengthscale, but quickly becomes inaccurate as the distances become larger. For Jaccard distances we need to calculate the reciprocal of $|A \cup B|$. In the dataset we are using, the value of $|A \cup B|$ lies between 35 and 105 about 99.9\% of the time. We approximate the reciprocal function with a degree 6 Taylor expansion about 70, which gives a fairly close approximation between these values.

These approximations, combined with the calculation of distances and multiplication of matrices, result in a degree 27 polynomial for the whole GP algorithm when using Hamming distances, for which we used a modulus chain of length 13, a degree 255 polynomial when using Jaccard distances, for which we used a modulus chain of length 22, and a degree 15 polynomial when using Euclidean distances, for which we used a modulus chain of length 11. The inaccuracy caused by approximating with these polynomials turns out to be so large that the predictions are essentially random, so these provide nothing more than a lower bound on the degree (and modulus chain length) required for accurate predictions. Attaining a reasonable level of accuracy for predictions would have required much larger degree polynomials, which we were unable to implement due to the exploding computational time.

We record the time taken between the test data being encrypted and the result being decrypted, and also the accuracy of the calculation as compared to a plaintext version of the algorithm. In both cases we calculate the inverse of the covariance matrix $K$ before we start timing, as this does not depend on the client's data. We record times for the rest of the process using Hamming, Jaccard and Euclidean distances with 5 training data points and 5 test data points, and in the modular approach we also record the time taken for each step. When using Hamming distances we use a lengthscale of 11.0315, for Jaccard distances we use a lengthscale of 0.1175 and for Euclidean distances we use a  lengthscale of 3.8058. These lengthscales were found by optimising the log marginal likelihood of the data given the lengthscale, and thus represent realistic lengthscales for data of this kind. All implementations use a security parameter of 80. Real world applications would likely require a larger security parameter than this, but for the sake of comparing methods we chose a low security level to make the runtime and memory footprint of the non-modular approach more managable.
\begin{table}[hbt]
\centering
\begin{tabular}{|r|c|c|} \cline{2-3}
\multicolumn{1}{c|}{\textbf{Time (s)}} & Modular & Non-modular \\
\hline
Hamming distance & 32 & - \\
\hline
Jaccard distance & 80 & - \\
\hline
Euclidean distance & 39 & - \\
\hline
Kernel function & 0 & - \\
\hline
Matrix multiplication & 6 & - \\
\hlineB{3}
\multicolumn{1}{V{3}rV{1.5}}{
Total (Hamming)
}&
\multicolumn{1}{V{1.5}cV{1.5}}{
38
}&
\multicolumn{1}{V{1.5}cV{3}}{
2684
}\\
\hlineB{3}
\multicolumn{1}{V{3}rV{1.5}}{
Total (Jaccard)
}&
\multicolumn{1}{V{1.5}cV{1.5}}{
86
}&
\multicolumn{1}{V{1.5}cV{3}}{
48677
}\\
\hlineB{3}
\multicolumn{1}{V{3}rV{1.5}}{
Total (Euclidean)
}&
\multicolumn{1}{V{1.5}cV{1.5}}{
45
}&
\multicolumn{1}{V{1.5}cV{3}}{
411
}\\
\hlineB{3}
\end{tabular}

\begin{tabular}{|r|c|c|} 
\multicolumn{3}{c}{ } \\
\cline{2-3}
\multicolumn{1}{c|}{\textbf{Accuracy}} & Modular & Non-modular \\
\hlineB{3}
\multicolumn{1}{V{3}rV{1.5}}{
Total (Hamming)
}&
\multicolumn{1}{V{1.5}cV{1.5}}{
\textpm $\:$ 0.00004 \%
}&
\multicolumn{1}{V{1.5}cV{3}}{
\textpm $\:$ 587863 \%
}\\
\hlineB{3}
\multicolumn{1}{V{3}rV{1.5}}{
Total (Jaccard)
}&
\multicolumn{1}{V{1.5}cV{1.5}}{
\textpm $\:$ 0.00286 \%
}&
\multicolumn{1}{V{1.5}cV{3}}{
\textpm $\:$ 145803473 \%
}\\
\hlineB{3}
\multicolumn{1}{V{3}rV{1.5}}{
Total (Euclidean)
}&
\multicolumn{1}{V{1.5}cV{1.5}}{
\textpm $\:$ 1.754 \%
}&
\multicolumn{1}{V{1.5}cV{3}}{
\textpm $\:$ 59.206 \%
}\\
\hlineB{3}
\end{tabular}
\caption{A comparison of the computation times and accuracy for regression using a Gaussian process on 5 training data points and 5 test data points for modular and non-modular approaches. Accuracy is recorded as the average difference between the result and the same result taken from a plaintext implementation of the algorithm, expressed as a percentage of the plaintext result.}\label{table}
\end{table}

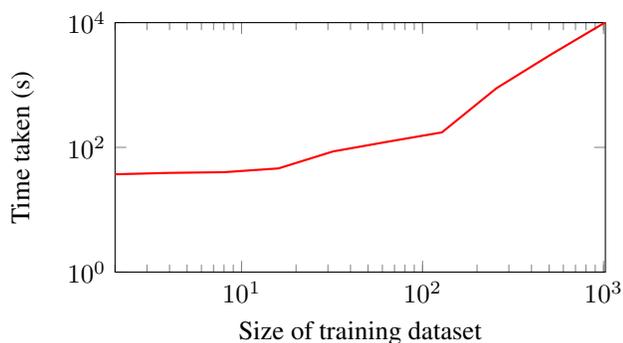
\begin{figure}[ht]
\begin{tikzpicture}
  \begin{axis}[ 
    xmode = log,
    ymode = log,
  	xmin = 2,
  	xmax = 1024,
  	ymin = 1,
  	ymax = 10000,
  	width = 8.1cm,
  	height = 4.9cm,
  	xlabel = Size of training dataset,
  	ylabel = Time taken (s),
  ] 
\addplot [red, thick] table {
    x 	 y
    2   37
    4   39
    8   40
    16  46
    32  86
    64  123
    128 174
    256 887
    512 3079
    1024 10137
    }; 
  \end{axis}
\end{tikzpicture}
\caption{ A graph showing how execution time of a GP scales with the size of the training dataset. All timings are for the whole GP regression workflow using Hamming distance with a test dataset of size 10.}
\label{timing}
\end{figure}

We tested on a POWER8+ processor with Infiniband interconnect, using 10 threads. Table \ref{table} shows that the modular approach is both faster and more accurate than the non-modular approach, with speedups of up to 566x even when the non-modular approach's approximation was so rough as to be meaningless. The level of inaccuracy of the non-modular approach would render the algorithm useless for any real world application. Thus it would be necessary to use a higher-degree approximation, which would increase the time taken even further. Figure \ref{timing} shows how the size of the dataset affects speed under the modular approach.

\section{Summary}\label{summary}

In this paper we have demonstrated how FHE can be effectively applied to Gaussian process regression, a popular lazy learning technique. In order to mitigate the large computational cost of operations on encrypted data within FHE, with respect to their plaintext equivalents, we break the Gaussian process methodology into stages, not all of which are required to be performed under encryption. This can be used to offer efficient, accurate prediction against a machine learning model as a service without the need to establish trust between client and service provider. The ability to make efficient, accurate predictions on encrypted data would be of great value in fields such as healthcare or banking, where machine learning can be used for tasks such as drug discovery or fraud detection, but privacy of data is essential.

We have shown that this modular approach offers significant improvements when compared to an approach with a single encryption stage. In a single stage approach, parts of the Gaussian process algorithm would need to be approximated, leading to a loss of accuracy and an increase in computational cost. The modular approach removes the need to approximate these parts, thus making the calculation significantly more accurate, as well as faster. We believe that this represents the first time a Gaussian process workflow has been performed under encryption, possibly due to the impracticalities of implementing such an algorithm in a non-modular fashion.

\section*{Acknowledgements}
This work was supported by the STFC Hartree Centre’s Innovation Return on Research programme, funded by the Department for Business, Energy \& Industrial Strategy.

\bibliographystyle{aaai}
\bibliography{bibliography}
\end{document}